\documentstyle[aps,epsfig]{revtex}

\newcommand{\mt}[1]{\mbox{\tiny #1}}
\newcommand{\bv}{\overline{V}}
\newcommand{\barq}{\langle\bar{q}q\rangle}
\newcommand{\abarq}{\langle\bar{q}i\gamma_5q\rangle}
\newcommand{\pe}{p_{\tiny{E}}}
\newcommand{\qe}{q_{\tiny{E}}}
\newcommand{\dtmu}{{\cal D}_{T,\mu}(p)}
\newcommand{\stmu}{\Sigma_{T,\mu}(p)}
\newcommand{\stmusq}{\Sigma_{T,\mu}^2(p)}
\newcommand{\ptmu}{\Sigma_{5;T,\mu}(p)}
\newcommand{\ptmusq}{\Sigma_{5;T,\mu}^2(p)}
\newcommand{\gev}{\mbox{\small GeV}}

\begin{document}
\tightenlines
\draft

\title{Current quark mass effects on chiral phase transition\\
of QCD in the improved ladder approximation}

\author{O.  Kiriyama \thanks{Email address: kiriyama@rcnp.osaka-u.ac.jp}}
\address{Research Center for Nuclear Physics, Osaka University, Ibaraki 567--0047, Japan}
\author{M.  Maruyama
\thanks{Email address: maruyama@nucl.phys.tohoku.ac.jp} and 
F.  Takagi\thanks{Email address: takagi@nucl.phys.tohoku.ac.jp}}
\address{Department of Physics, Tohoku University, Sendai 980--8578, Japan}
\maketitle

\begin{abstract}
Current quark mass effects on the chiral phase transition of QCD 
is studied in the improved ladder approximation. 
An infrared behavior of the gluon 
propagator is modified in terms of an effective running coupling. 
The analysis is based on a composite operator formalism 
and a variational approach. 
We use the Schwinger-Dyson equation to give a ``normalization 
condition'' for the Cornwall-Jackiw-Tomboulis effective potential 
and to isolate the ultraviolet divergence which appears in an 
expression for the quark-antiquark condensate. 
We study the current quark mass effects on the order parameter 
at zero temperature and density. 
We then calculate the effective potential at finite temperature and density 
and investigate the current quark mass effects on the chiral phase transition. 
We find a smooth crossover for $T>0$, $\mu=0$ 
and a first-order phase transition for 
$\mu>0$, $T=0$. Critical exponents are also studied and 
our model gives the classical mean-field values. 
We also study the temperature dependence of masses of scalar 
and pseudoscalar bosons. A critical end point in the $T$-$\mu$ plane
is found at $T \sim 100$ MeV, $\mu \sim 300$ MeV.
\end{abstract}


\section{INTRODUCTION}
Recently there has been great interest in studying the phase structure of 
quantum chromodynamics (QCD). We expect that at sufficiently high temperature 
and/or density the QCD vacuum changes into a chirally symmetric/
deconfinement phase \cite{MCLERRAN}, a color superconducting phase 
\cite{BAILIN,IWASAKI,WILCZEK} or a color-flavor locked phase \cite{SCHAFER}. 
They may be realized in high-energy heavy-ion collisions at 
the BNL Relativistic Heavy Ion Collider and CERN Large Hadron Collider. 
These phase transitions are important also in the physics of neutron stars 
and the early universe. 

The phase structure of QCD has been studied by various methods. 
At finite temperature the lattice simulation is powerful 
for studying the phase structure of QCD. 
On the other hand, the so-called QCD-like theories, 
one category of the effective theories of QCD, 
are still useful for studying the chiral phase structure at high temperature 
and/or density [6--15]. QCD in the weak coupling limit is utilized 
to study the color superconductivity [2--5]. 
However, most studies have been done in the zero external field limit; i.e., 
in the chiral limit for the case of the dynamical chiral symmetry breaking. 
It is, then, desirable to study a more realistic situation where the 
chiral symmetry is explicitly broken by a current quark mass. 
In this paper we neglect the quark pairing and study the current quark mass 
effects on the chiral phase transition between 
$SU(N_c) \times SU(N_f)_L \times SU(N_f)_R$ 
and  $SU(N_c) \times SU(N_f)_{L+R}$. 
In order to study nonperturbative phenomena such as 
the dynamical chiral symmetry breaking, the Schwinger-Dyson equation (SDE) 
or the effective potential for a composite operator has been widely used. 
However, in order to find the true vacuum it is necessary to calculate 
the effective potential. 
Furthermore, in the studies of the SDE, 
it was known that there is a difficulty 
in removing a perturbative contribution which is quadratically divergent 
from the order parameter of chiral symmetry \cite{MIRANSKY,KUSAKA}.

Our analysis starts from the Cornwall-Jackiw-Tomboulis (CJT) 
effective action for a composite operator \cite{CJT} 
and QCD in the improved ladder (rainbow) approximation [19--21]. 
The improved ladder approximation is 
the simplest nonperturbative truncation scheme which is consistent 
with the renormalization group. At zero temperature and density, 
it reproduces the physical quantities 
which are insensitive to the model parameter \cite{AOKI}. 
Therefore, we expect that it is valid also 
at finite temperature and/or density. 
We use the SDE to modify the CJT effective potential. 
The result is used to give a ``normalization condition'' for 
the effective potential \cite{BARDUCCI88} and to isolate 
the ultraviolet divergence 
which appears in an expression for quark-antiquark condensate. 
The divergence depends solely on the qurak mass and is independent of the 
condensate.

This paper is organized as follows. In Sec. II we derive 
the modified form of the CJT effective potential 
for quark propagator at zero temperature and density. 
The critical value of the coupling constant and 
the critical number of quark flavors are examined for 
comparison with the results by other authors. 
We then extend the modified effective potential 
to finite temperature and density. 
In Sec. III we first determine the value of $\Lambda_{\mt{QCD}}$ 
by a condition $f_{\pi}=93$ MeV at $T=\mu=0$ and in the chiral limit. 
Then we calculate the effective potential numerically and investigate 
the phase structure for nonzero $T,\mu$ and nonzero current quark mass. 
We also examine the critical exponents, 
the temperature dependence of masses of the scalar and pseudoscalar
bosons and the critical end point. 
Section IV is devoted to the summary and discussion. 
We fix the mass scale by the condition $\Lambda_{\mt{QCD}}=1$ except 
for numerical calculations.
   
\section{EFFECTIVE POTENTIAL FOR THE QUARK PROPAGATOR}
\subsection{CJT effective potential at zero temperature and density}

At zero temperature and zero density, the CJT effective potential 
for QCD in the improved ladder approximation is expressed 
as a functional of $S(p)$ the quark full propagator \cite{HIG83}:
\begin{eqnarray}
V[S]&=&V_1[S]+V_2[S],\\
V_1[S]&=&\int\frac{d^4p}{(2\pi)^4i}~
\mbox{Tr}\left\{\ln [S_0^{-1}(p)S(p)]-S_0^{-1}(p)S(p)+1\right\},
\label{eqn:vg1}\\
V_2[S]&=&-\frac{i}{2}C_2\int\int\frac{d^4p}{(2\pi)^4i}~
\frac{d^4q}{(2\pi)^4i}~\bar{g}^2(p,q)\mbox{Tr}
\left[\gamma_{\mu}S(p)\gamma_{\nu}S(q)\right]D^{\mu\nu}(p-q),\label{eqn:vg2}
\end{eqnarray}
where  $C_2=(N_c^2-1)/(2N_c)$ is the 
quadratic Casimir operator for the color $SU(N_c)$ group, 
$S_0(p)$ is the bare quark propagator, 
$\bar{g}^2(p,q)$ is the QCD running coupling of one-loop order, 
$D^{\mu\nu}(p)$ is the gluon propagator which is diagonal in the color space 
and, ``Tr'' refers to Dirac, flavor, and color matrices. 
The two-loop potential $V_2$ is given by the vacuum graph of the fermion 
one-loop diagram with one-gluon exchange.

After Wick rotation, we use the following approximation
\begin{eqnarray}
\bar{g}^2(\pe,\qe)=\theta(\pe-\qe)\bar{g}^2(\pe)+\theta(\qe-\pe)\bar{g}^2(\qe).
\end{eqnarray}
This approximation is occasionally called 
Higashijima-Miransky approximation \cite{HIG83,MIR}. 
In this approximation and in the Landau gauge, renormalization of the 
quark wave function is unnecessary and 
the CJT effective potential is expressed 
in terms of $\Sigma(\pe)$, the dynamical mass function of a quark as follows
\begin{eqnarray}
V[\Sigma(\pe)]&=&V_1[\Sigma(\pe)]+V_2[\Sigma(\pe)],\\
V_1[\Sigma(\pe)]&=&-2\int^{\Lambda}\frac{d^4\pe}{(2\pi)^4}
~\ln\frac{\Sigma^2(\pe)+\pe^2}{m^2(\Lambda)+\pe^2}\nonumber\\
&&+4\int^{\Lambda}\frac{d^4\pe}{(2\pi)^4}
~\frac{\Sigma(\pe)[\Sigma(\pe)-m(\Lambda)]}
{\Sigma^2(\pe)+\pe^2},\label{eqn:app5}\\
V_2[\Sigma(\pe)]&=&-6C_2\int^{\Lambda}\int^{\Lambda}\frac{d^4\pe}{(2\pi)^4}
\frac{d^4\qe}{(2\pi)^4}~\frac{\bar{g}^2(\pe,\qe)}{(\pe-\qe)^2}\nonumber\\
&&\times\frac{\Sigma(\pe)}{\Sigma^2(\pe)+\pe^2}
\frac{\Sigma(\qe)}{\Sigma^2(\qe)+\qe^2},
\label{eqn:app6}
\end{eqnarray}
where an overall factor 
(the number of light quarks times the number of colors) is 
omitted and $m(\Lambda)$ is the bare quark mass.
In the above equations we temporarily introduced 
the ultraviolet cutoff $\Lambda$ 
in order to make the bare quark mass well defined.

The extremum condition for $V$ with respect to the variation of $\Sigma(\pe)$ 
leads to the following SDE for the quark self-energy
\begin{eqnarray}
\Sigma(\pe)=m(\Lambda)+3C_2\int^{\Lambda}\frac{d^4\qe}{(2\pi)^4}~
\frac{\bar{g}^2(\pe,\qe)}{(\pe-\qe)^2}~\frac{\Sigma(\qe)}{\Sigma^2(\qe)+\qe^2}.
\label{eqn:sd-1}
\end{eqnarray}
In the Higashijima-Miransky approximation, since the argument of the running 
coupling has no angle dependence, we first perform the angle integration 
using the result:
\begin{eqnarray}
\int\frac{d\Omega_4}{(\pe-\qe)^2}=\theta(\pe-\qe)\frac{1}{\pe^2}
                                  +\theta(\qe-\pe)\frac{1}{\qe^2},
\end{eqnarray}
where
\begin{eqnarray}
\int d\Omega_4=\frac{1}{2\pi^2}\int_0^{\pi}d\psi\sin^2\psi
	       \int_0^{\pi}d\theta\sin\theta
	       \int_0^{2\pi}d\phi.
\end{eqnarray}
Then Eq. (\ref{eqn:sd-1}) is reduced 
to the following differential equation \cite{MIRANSKY,HIG91}:
\begin{eqnarray}
\frac{\Sigma(\pe)}{\Sigma^2(\pe)+\pe^2}=\frac{(4\pi)^2}{3C_2}
\frac{d}{\pe^2d\pe^2}\left(\frac{1}{\Delta(\pe)}
\frac{d\Sigma(\pe)}{d\pe^2}\right),
\label{eqn:dsd}
\end{eqnarray}
with the two boundary conditions
\begin{eqnarray}
\frac{1}{\Delta(\pe)}
\frac{d\Sigma(\pe)}{d\pe^2}~\Bigg|_{\pe=0}&=&0,\label{eqn:app14}\\
\Sigma(\pe)-\frac{{\cal D}(\pe)}{\Delta(\pe)}
\frac{d\Sigma(\pe)}{d\pe^2}~\Bigg|_{\pe=\Lambda}&=&m(\Lambda),\label{eqn:app15}
\end{eqnarray}
where the functions
\begin{eqnarray}
{\cal D}(\pe)=\frac{\bar{g}^2(\pe)}{\pe^2},
\end{eqnarray}
and
\begin{eqnarray}
\Delta(\pe)=\frac{d}{d\pe^2}{\cal D}(\pe),
\end{eqnarray}
are introduced. 

Substituting Eqs. (\ref{eqn:sd-1}) and (\ref{eqn:dsd}) 
into Eqs. (\ref{eqn:app5}) and (\ref{eqn:app6}), we obtain
\begin{eqnarray}
V[\Sigma(\pe)]&=&-2\int^{\Lambda}\frac{d^4\pe}{(2\pi)^4}~
\ln\frac{\Sigma^2(\pe)+\pe^2}{m^2(\Lambda)+\pe^2}\nonumber\\
&&+2\int^{\Lambda}\frac{d^4\pe}{(2\pi)^4}
~\frac{\Sigma(\pe)[\Sigma(\pe)-m(\Lambda)]}
{\Sigma^2(\pe)+\pe^2}\nonumber\\
&=&-2\int^{\Lambda}\frac{d^4\pe}{(2\pi)^4}~\ln\frac{\Sigma^2(\pe)+\pe^2}
{m^2(\Lambda)+\pe^2}\nonumber\\
&&+\frac{2(4\pi)^2}{3C_2}\int^{\Lambda}\frac{d^4\pe}{(2\pi)^4}~
\left[\Sigma(\pe)-m(\Lambda)\right]
\frac{d}{\pe^2d\pe^2}
\left(\frac{1}{\Delta(\pe)}\frac{d\Sigma(\pe)}{d\pe^2}\right)\nonumber\\
&=&-2\int^{\Lambda}\frac{d^4\pe}{(2\pi)^4}~
\ln\frac{\Sigma^2(\pe)+\pe^2}{m^2(\Lambda)+\pe^2}\nonumber\\
&&-\frac{2}{3C_2}\int^{\Lambda^2}d\pe^2~\frac{1}{\Delta(\pe)}
\left(\frac{d}{d\pe^2}\Sigma(\pe)\right)^2+V_S,\label{eqn:app12}
\end{eqnarray}
where we used a partial integration in the last line 
and
\begin{eqnarray}
V_S&=&F(\Lambda)-F(0),\label{eqn:NC}\\
F(\pe)&=&\frac{2}{3C_2}\left[\Sigma(\pe)-m(\Lambda)\right]
\frac{1}{\Delta(\pe)}\frac{d\Sigma(\pe)}{d\pe^2}.\label{eqn:app13}
\end{eqnarray}

Hereafter, we consider the effective potential 
in the continuum limit $(\Lambda\to\infty)$. 
Let us begin by evaluating $F(\Lambda)$ using the running coupling
\begin{eqnarray}
\bar{g}^2(\pe)=\frac{2\pi^2a}{\ln \pe^2}~~,~~a\equiv\frac{24}{11N_c-2n_f},
\label{eqn:asym1}
\end{eqnarray}
and the corresponding asymptotic form of the mass function
\begin{eqnarray}
\Sigma(\pe)\rightarrow 
m(\Lambda)\left(\frac{\ln \pe^2}{\ln \Lambda^2}\right)^{-a/2}
+\frac{\sigma}{\pe^2}(\ln \pe^2)^{a/2-1},
\label{eqn:asym2}
\end{eqnarray}
where $n_f$ is the number of quark flavors 
which contributes to the running coupling. 
Throughout this paper, with the exception of discussion on 
a critical number of massless flavors, we put $N_c=n_f=3$, namely, $a=8/9$. 
As we will show below, the parameter $\sigma$ is related to 
the order parameter of chiral symmetry, a.k.a. the quark condensate. 
Note that Eq. (\ref{eqn:asym2}) is to be understood in the sense that 
for exact chiral symmetry the first term on the RHS is zero and the second 
term is the dominant one, while in the presence of explicit chiral symmetry 
breaking the first term gives the dominant asymptotic behavior. 
With the explicit chiral symmetry breaking, 
there will be many terms which are 
suppressed by powers of $\ln(\pe^2)$ compared to the first term, but 
dominate the second term as $\pe^2\to\infty$ \cite{MIRANSKY}.

Using Eqs. (\ref{eqn:asym1}) and (\ref{eqn:asym2}), we obtain
\begin{eqnarray}
F(\Lambda)&=&\frac{2}{3C_2}\cdot\frac{\sigma^2}{\Lambda^2}~
             (\ln \Lambda^2)^{a/2-1}
             ~\frac{\Lambda^4(\ln \Lambda^2)^2}{\ln \Lambda^2+1}\cdot
             \frac{-1}{2\pi^2a}\nonumber\\
&&\times\left[-\frac{a}{2}\cdot\frac{m(\Lambda)}{\Lambda^2\ln\Lambda^2}
-\frac{\sigma}{\Lambda^4}(\ln\Lambda^2)^{a/2-2}(\ln\Lambda^2+1-a/2)\right]
\nonumber\\
&\simeq&\frac{2}{6\pi^2aC_2}\Lambda^2(\ln\Lambda^2)^{a/2}
\left[\frac{a}{2}\cdot\frac{m(\Lambda)}{\Lambda^2\ln\Lambda^2}
+\frac{\sigma}{\Lambda^4}(\ln\Lambda^2)^{a/2-1}\right].\label{eqn:app19}
\end{eqnarray}
We note that $m(\Lambda)$ is the bare quark mass 
defined at the scale $\Lambda$, as mentioned before, 
and the factor $(\ln\Lambda^2)^{a/2}m(\Lambda)$ in Eq. (\ref{eqn:app19}) 
is equal to $(\ln\kappa^2)^{a/2}m_R(\kappa)$ which is cutoff independent.
\footnote{In this paper, we use the mass-independent renormalization scheme 
\cite{WEINBERG,THOOFT}. 
In this scheme, all the renormalization constants are fixed 
by massless theory. The bare quark mass $m(\Lambda)$ and the renormalized 
mass $m_R(\kappa)$ at the renormalization point $\kappa$ are related as
\begin{eqnarray}
m(\Lambda)=Z_S^{-1}(\Lambda,\kappa)m_R(\kappa)\nonumber
\end{eqnarray}
where $Z_S^{-1}(\Lambda,\kappa)$ is the renormalization constant 
for the composite operator $\bar{q}q$:
\begin{eqnarray}
(\bar{q}q)_R&=&Z_S^{-1}\cdot(\bar{q}q)_{\Lambda},\nonumber\\
Z_S^{-1}(\Lambda,\kappa)&=&
\left(\frac{\ln \kappa^2}{\ln \Lambda^2}\right)^{a/2}.
\nonumber
\end{eqnarray}
} Hence, $F(\Lambda)$ vanishes 
in the continuum limit, provided $\sigma$ remains finite 
in this limit. As concerns $F(0)$, it turns out to be
\begin{eqnarray}
F(0)&=&\frac{2}{3C_2}\left[\Sigma(\pe)-m(\Lambda)\right]\nonumber\\
&&\times\frac{(\pe^2)^2}
        {\pe^2d\bar{g}^2(\pe)/d\pe^2-\bar{g}^2(\pe)}\frac{d\Sigma(\pe)}{d\pe^2}
        ~\Bigg|_{\pe=0}.
\end{eqnarray}
Since we will introduce an infrared finite running coupling and 
mass function in Eqs. (\ref{eqn:ec}) and (\ref{eqn:mass}), we can set $F(0)=0$.

After all, in the continuum limit, we get $V_S=0$ and the modified 
version of the CJT effective potential becomes
\begin{eqnarray}
V[\Sigma(\pe)]&=&-2\int\frac{d^4\pe}{(2\pi)^4}~
\ln\frac{\Sigma^2(\pe)+\pe^2}{\pe^2}\nonumber\\
&&-\frac{2}{3C_2}\int d\pe^2~\frac{1}{\Delta(\pe)}
\left(\frac{d}{d\pe^2}\Sigma(\pe)\right)^2.
\label{eqn:vcjt}
\end{eqnarray}

A few comments are in order.

(1) The extremum condition for $V[\Sigma(\pe)]$ in
Eq. (\ref{eqn:vcjt}) with respect 
to $\Sigma(\pe)$ leads to Eq. (\ref{eqn:dsd}) which is equivalent to 
the original equation (\ref{eqn:sd-1}) 
in the Higashijima-Miransky approximation 
apart from the two boundary conditions. 
We will take account of these conditions when we introduce the trial 
mass function.

(2) Even if chiral symmetry is explicitly broken, we do not use the 
condition adopted in Ref. \cite{BARDUCCI88}. Instead, our 
condition is $V_S=0$ (see Eq. (\ref{eqn:NC})).

The relation between $\sigma$ and the quark condensate is as follows. 
The renormalization group invariant vacuum expectation value of 
$\bar{q}q$ is given by
\begin{eqnarray}
\barq&=&\langle 0|(\bar{q}q)_R|0\rangle(\ln\kappa^2)^{-a/2}\nonumber\\
&=&\lim_{\Lambda\to\infty}
\langle 0|(\bar{q}q)_{\Lambda}|0\rangle(\ln\Lambda^2)^{-a/2}\nonumber\\
&=&\lim_{\Lambda\to\infty}G(\Lambda),\label{eqn:dbarq}
\end{eqnarray}
where $G(\Lambda)$ is defined as 
\begin{eqnarray}
G(\Lambda)=Z^{-1}(\Lambda)(-4N_c)\int^{\Lambda}\frac{d^4\pe}{(2\pi)^4}~
\frac{\Sigma(\pe)}{\Sigma^2(\pe)+\pe^2},\label{eqn:glambda}
\end{eqnarray}
with
\begin{eqnarray}
Z(\Lambda)=(\ln \Lambda^2)^{a/2}.
\end{eqnarray}
Using Eq. (\ref{eqn:dsd}), $G(\Lambda)$ is rewritten as
\begin{eqnarray}
G(\Lambda)&=&-Z^{-1}(\Lambda)\frac{4N_c}{3C_2}
\int_0^{\Lambda^2}d\pe^2\frac{d}{d\pe^2}\left(\frac{1}{\Delta(\pe)}
\frac{d\Sigma(\pe)}{d\pe^2}\right).
\end{eqnarray}
This expression is convenient to isolate the ultraviolet divergence as 
the RHS is linear in $\Sigma(\pe)$. 
Using Eqs. (\ref{eqn:asym1}) and (\ref{eqn:asym2}), as can be verified by 
direct calculation, $G(\Lambda)$ is obtained as
\begin{eqnarray}
G(\Lambda)=m(\Lambda) I_p(\Lambda)+\sigma I_n(\Lambda),\label{eqn:condensate}
\end{eqnarray}
where the lower bound in the definite integral
automatically vanishes by the same argument as given for $F(0)$ and 
a perturbative contribution $I_p(\Lambda)$ and 
a nonperturbative one $I_n(\Lambda)$ are given as follows:
\begin{eqnarray}
I_p(\Lambda)&=&-\frac{N_c}{3\pi^2C_2}\Lambda^2(\ln\Lambda^2)^{-a/2},\\
I_n(\Lambda)&=&-\frac{2N_c}{3\pi^2aC_2}.
\end{eqnarray}
As discussed in Ref. \cite{MIRANSKY,KUSAKA}, the perturbative contribution is 
quadratically divergent. We are interested in 
the purely nonpertubetive effect; 
therefore, we simply subtract the perturbative effect from 
Eq. (\ref{eqn:condensate}) and re-define the quark condensate as 
\begin{eqnarray}
\barq&\equiv&\sigma I_n(\Lambda)\nonumber\\
&=&-\frac{3}{2\pi^2a}\sigma.\label{eqn:cc}
\end{eqnarray}
The relation between $\sigma$ and $\barq$ is the same as in 
the massless theory. We do not use 
Eq. (\ref{eqn:dbarq}) but Eq. (\ref{eqn:cc}) to extract $\barq$ from 
$\sigma_{min}$, the location of a minimum of the effective potential. 
Consequently, the asymptotic behavior of the mass function 
in the continuum limit takes the following form
\begin{eqnarray}
\Sigma(\pe) \rightarrow m_R(\ln \pe^2)^{-a/2}
-\barq\frac{2\pi^2a}{3\pe^2}(\ln \pe^2)^{a/2-1},\label{eqn:asym3}
\end{eqnarray}
where
\begin{eqnarray}
m_R=m_R(\kappa)(\ln \kappa^2)^{a/2}
=\lim_{\Lambda\to\infty}m(\Lambda)(\ln \Lambda^2)^{a/2}\label{eqn:rgmass}
\end{eqnarray}
is the renormalization group invariant quark mass. 
This asymptotic behavior is completely coincident with that 
obtained by using the operator product expansion and the 
renormalization group equation \cite{POLITZER}. 
Notice that in these arguments there 
are no ambiguities in separating the perturbative and 
the nonperturbative contribution in $\barq$.

Now we are in a position to introduce a modified running coupling and a 
trial mass function. We use the following 
modified running coupling \cite{HIG91}
\begin{eqnarray}
\bar{g}^2(\pe)=\frac{2\pi^2a}{\ln (\pe^2+p_R^2)},\label{eqn:ec}
\end{eqnarray}
where $p_R$ is a parameter to regularize the divergence of the
 QCD running coupling at $\pe=1(\Lambda_{\mt{QCD}})$. 
This modified running coupling approximately develops according to the 
QCD renormalization group equation of one-loop order at large $\pe^2$, 
while it smoothly approaches a constant as $\pe^2$ decreases. 

Corresponding to the above running coupling, the SDE 
with the two boundary conditions suggests 
the following trial mass function
\begin{eqnarray}
\Sigma(\pe)=m_R[\ln (\pe^2+p_R^2)]^{-a/2}
+\frac{\sigma}{\pe^2+p_R^2}[\ln (\pe^2+p_R^2)]^{a/2-1}.\label{eqn:mass}
\end{eqnarray}
The trial mass function is infrared 
finite, and moreover, it has the same asymptotics as in Eq. (\ref{eqn:asym3}).
 
When we also consider the pseudoscalar degree of freedom 
in the effective potential, 
we expand the full quark propagator in Minkowski space as follows
\begin{eqnarray}
iS^{-1}(p)=\gamma\cdot p-\Sigma(p)-i\gamma_5\Sigma_5(p).
\end{eqnarray}
An effective potential for this case is obtained straightforwardly as 
\begin{eqnarray}
V[\Sigma(\pe),\Sigma_5(\pe)]&=&-2\int\frac{d^4\pe}{(2\pi)^4}~
\ln\frac{\Sigma^2(\pe)+\Sigma_5^2(\pe)+\pe^2}{\pe^2}\nonumber\\
&&-\frac{2}{3C_2}\int d\pe^2~\frac{1}{\Delta(\pe)}
\left[\left(\frac{d}{d\pe^2}\Sigma(\pe)\right)^2
+\left(\frac{d}{d\pe^2}\Sigma_5(\pe)\right)^2\right].
\label{eqn:vcjt2}
\end{eqnarray}
The mass function $\Sigma_5(\pe)$ satisfies 
the SDE which has the form as in Eq. (\ref{eqn:dsd}) 
where the bare quark mass is set zero. 
Therefore the trial mass function for $\Sigma_5(\pe)$ should be
\begin{eqnarray}
\Sigma_5(\pe)=\frac{\sigma_5}{\pe^2+p_R^2}
[\ln (\pe^2+p_R^2)]^{a/2-1},\label{eqn:mass2}
\end{eqnarray}
where $\sigma_5=2\pi^2a\abarq/3$. 
However, except when we consider a mass of pseudoscalar boson 
we can put $\Sigma_5(\pe)=0$ without loss of generality. 

Substituting Eqs. (\ref{eqn:ec}) and (\ref{eqn:mass}) 
into Eq. (\ref{eqn:vcjt}), we have the following expression for the 
effective potential
\begin{eqnarray}
V(\sigma;m_R)&=&-\frac{2}{(4\pi)^2}\int_{t_R}^{\infty}dt~e^t
(e^t-e^{t_R})\ln
\frac{e^t-e^{t_R}+(m_Rt^{-a/2}+\sigma t^{a/2-1}e^{-t})^2}
{e^t-e^{t_R}+m_R^2t^{-a}}
\nonumber\\
&&+\frac{m_R\sigma}{4\pi^2}\int_{t_R}^{\infty}dt~
\frac{e^{-t}(e^t-e^{t_R})^2(t+1-a/2)}{t(e^t-e^{t_R}+te^t)}\nonumber\\
&&+\frac{9}{32\pi^2}\sigma^2\int_{t_R}^{\infty}dt~
\frac{(1-e^{t_R-t})^2(t+1-a/2)^2}{t^{2-a}(e^t-e^{t_R}+te^t)},\label{eqn:vsm2}
\end{eqnarray}
where we made the change of the integration variable
\begin{eqnarray}
t=\ln (\pe^2+p_R^2)~~,~~t_R=\ln (p_R^2),
\end{eqnarray}
and subtracted the $\sigma$ independent term 
which is exponentially divergent in $t$. 
The infrared regularization parameter $t_R$ specifies 
the coupling constant in the low energy region. 
Indeed, $\bar{g}^2(0)=2\pi^2a/t_R$ is proportional to the inverse of $t_R$. 
We note that in the first two terms in Eq. (\ref{eqn:vsm2}) 
there are logarithmic divergences in $t$ which cancel out each other. 
This is because the fact that we take into 
account the correct renormalization group effects 
in $\bar{g}^2(\pe)$ and in $\Sigma(\pe)$ \cite{BARDUCCI88}.

Before we turn to the effective potential at finite temperature 
and density, we examine some numerical consequences 
of Eq. (\ref{eqn:vsm2}) and make a comparison with the results 
by other authors. 
Figure 1 shows the $t_R$ dependence of $\barq$ in the chiral limit. 
Chiral symmetry is broken for relatively small $t_R$ whereas it is restored 
for large $t_R$. 
Notice that $\barq$ is stable under the change of $t_R$ if $t_R<0.4$ 
and the critical value of $t_R$ is $(t_R)_{critical}\simeq 0.56$. 
This critical value of $t_R$ is 
about one third of that obtained in Ref. \cite{HIG91} 
in which the SDE with the same running coupling has been solved. In our model,
\begin{eqnarray}
(\alpha_s)_{critical} &=& \frac{2\pi^2a}{(t_R)_{critical}}\nonumber\\
& \simeq & 3\cdot\frac{\pi}{3C_2},
\end{eqnarray}
where $\pi/(3C_2)$ is the critical value obtained 
from the SDE analysis in Ref. \cite{HIG91} and also 
in quenched $\mbox{QED}_4$ with 
the replacement $\alpha$ by $C_2\alpha$ \cite{MIRANSKY}. 
We note that, in Ref. \cite{HIG83,HIG91}, 
it has been shown that the $(\alpha_s)_{critical}$ 
obtained by using the CJT effective potential 
(\ref{eqn:app5}) and (\ref{eqn:app6}) is about two times of $\pi/(3C_2)$. 
These differences may arise from the fact that,
in the framework of the variational approach, 
we restricted the class of trial function and we used the modified 
form of the CJT effective potential.

Figure 2 shows the plot of the quark condensate $\barq$ versus 
quark mass $m_R$ analogous to the relation between 
the spontaneous magnetization of the ferro-magnetism 
and the external magnetic field. 
The curves show the cases $t_R=0.1$ and 
$t_R=0.8$. When $t_R$ is sufficiently small, i.e., 
the coupling constant in low energy region is sufficiently large, the 
quark condensate discontinuously changes its sign at $m_R=0$. This is a clear
 evidence of the dynamical chiral symmetry breaking. On the other hand, 
when $t_R$ is sufficiently large, the discontinuity of 
the quark condensate disappears.

The chiral phase, even at zero temperature and density, depends on the 
number of massless flavors. At some value of $n_f$ (less than $11N_c/2$), 
a phase transition to the chirally symmetric phase is expected. 
In Ref. \cite{APPELQUIST}, it has been argued that, in connection with the 
infrared fixed point in the two-loop $\beta$ function, 
there exists a critical number of fermions $n_f^{c}$, above which there is 
no chiral symmetry breaking. 
Figure 3 shows the $t_R$ dependence of $n_f^{c}$, 
where $n_f^{c}$ is considered as a continuous number. 
For $t_R=0.1$ our model suggests that 
$n_f^{c}=10$, and this result is consistent 
with lattice QCD results \cite{KOGUT} 
which gives $8 < n_f^{c} \le 12$ (see also Ref. \cite{APPELQUIST}).

\subsection{Effective potential at finite temperature and density}

In this subsection we discuss the effective potential 
at finite temperature and density. 
In order to calculate the effective potential at finite 
temperature and density we apply the imaginary time formalism \cite{JIK}
\begin{eqnarray}
\int\frac{dp_4}{2\pi}f(p_4) \to 
T\sum_{n=-\infty}^{\infty}f(\omega_n+i\mu),\label{eqn:itf}
\end{eqnarray}
where $\omega_n=(2n+1)\pi T$ $(n=0,\pm 1,\pm 2,\cdots)$ 
is the fermion Matsubara frequency 
and $\mu$ represents the quark chemical potential. 
In addition, we need to define the running coupling 
and the trial mass function at finite temperature and density. 
We adopt the following functions for $\dtmu$, $\stmu$ and $\ptmu$ by 
replacing $p_4$ in ${\cal D}(\pe)$, $\Sigma(\pe)$ and $\Sigma_5(\pe)$
with $\omega_n$:
\begin{eqnarray}
\dtmu&=&\frac{2\pi^2a}{\ln(\omega_n^2+|\vec{p}|^2+p_R^2)}~
\frac{1}{\omega_n^2+|\vec{p}|^2},\label{eqn:dp}\\
\stmu&=&m_R\left[\ln(\omega_n^2+|\vec{p}|^2+p_R^2)\right]^{-a/2}\nonumber\\
&&+\frac{\sigma}{\omega_n^2+|\vec{p}|^2+p_R^2}
\left[\ln(\omega_n^2+|\vec{p}|^2+p_R^2)\right]^{a/2-1}\label{eqn:sigma},\\
\ptmu&=&\frac{\sigma_5}{\omega_n^2+|\vec{p}|^2+p_R^2}
\left[\ln(\omega_n^2+|\vec{p}|^2+p_R^2)\right]^{a/2-1}\label{eqn:pi}.
\end{eqnarray}

In Eq. (\ref{eqn:dp}) we do not introduce the $\mu$ dependence 
in $\dtmu$. 
The gluon momentum squared is the most natural argument of the running 
coupling at zero temperature and density, in the light of the chiral 
Ward-Takahashi identity \cite{JAIN,KUGO}. 
Then it is reasonable to assume that $\dtmu$ 
does not depend on the quark chemical potential. 
In addition, the screening mass is not included 
in Eq. (\ref{eqn:dp}). We comment on this point in Sec. IV.

As concerns the mass function, we use the same function 
as Eqs. (\ref{eqn:mass}) and (\ref{eqn:mass2}) 
except that we replace $p_4$ with $\omega_n$. 
As already noted in Sec. II A, the quark wave function does not suffer the 
renormalization in the Landau gauge for $T=\mu=0$, while, the same does not 
hold for finite temperature and/or density. 
However, we assume that the wave function renormalization is not 
required even at finite temperature and/or density, for simplicity. 

Furthermore, we neglect the $T$-$\mu$ dependent 
terms in the quark and gluon propagators 
which arise from the perturbative expansion. 
We expect that the phase structure is not so affected by these approximations.

Using Eqs. (\ref{eqn:dp}), (\ref{eqn:sigma}) and (\ref{eqn:pi}), 
it is easy to write down 
the effective potential $V(\sigma,\sigma_5;m_R)$ (see Appendix). 
Even at finite temperature and/or density, 
we determine the value of $\barq$ through $\barq=-(3/2\pi^2a)\sigma_{min}$ 
where $\sigma_{min}$ is the location of 
the minimum of $V(\sigma,\sigma_5;m_R)$.

\section{CHIRAL PHASE TRANSITION AT HIGH TEMPERATURE AND 
DENSITY:NUMERICAL RESULTS}

In numerical calculation, as mentioned before, we put $N_c=n_f=3$. 
Furthermore, since it was known that 
the temperature and chemical potential dependence 
of the quantities such as $\barq$ and $f_{\pi}$ are stable under the change of 
the infrared regularization parameter \cite{TANIGUCHI}, 
in the first place, we fix $t_R=0.1$ and determine the value of 
$\Lambda_{\mt{QCD}}$ by the condition $f_{\pi}=93$ MeV at $T=\mu=0$ and 
$m_R=0$. In this case, the pion decay constant is 
approximately given by \cite{PS}
\begin{eqnarray}
f_{\pi}^2=4N_c\int\frac{d^4\pe}{(2\pi)^4}~
\frac{\Sigma(\pe)}{[\Sigma^2(\pe)+\pe^2]^2}
\left(\Sigma(\pe)-\frac{\pe^2}{2}\frac{d\Sigma(\pe)}{d\pe^2}\right),
\end{eqnarray}
and we have $\Lambda_{\mt{QCD}}=738$ MeV. 
Secondly, we assume the light quarks ($u$ and $d$) are degenerate in mass and 
take the quark mass evaluated at $\kappa=1$ GeV as 
\begin{eqnarray}
m_R(1\gev)=\frac{m_u(1\gev)+m_d(1\gev)}{2}=7{\rm MeV}.
\end{eqnarray}
With Eq. (\ref{eqn:rgmass}) the renormalization group 
invariant quark mass $m_R$ extracted from the above-mentioned value becomes
\begin{eqnarray}
m_R=7.6 \times 10^{-3}\Lambda_{\mt{QCD}}.
\end{eqnarray}

\subsection{$T \neq 0,\mu=0$ case}
The phase diagram in the chiral limit is shown in Fig. 4. We have found 
second-order phase transition, with the critical temperature $T_c=129$
MeV for $\mu=0$ \cite{KIRIYAMA2}.

Figure 5 shows the temperature dependence of the effective potential 
at $\mu=0$ for $m_R(1\gev)=7$ MeV. 
We can realize that $\sigma_{min}$ the minimum of the effective potential 
continuously approaches to the origin 
as temperature increases. 
Figure 6 shows the temperature dependence of $\sigma_{min}$ 
for the chiral limit and the case $m_R(1\gev)=7$ MeV. 
Below $T_c$, the two curves are almost the same 
apart from the small difference in their magnitude. 
On the other hand, above $T_c$, while $\sigma_{\min}$ remains nonzero 
for $m_R(1\gev)=7$ MeV, the one in the chiral limit 
vanishes. In other words, the temperature dependence of 
the order parameter is dominated 
by the dynamical symmetry breaking in low-$T$ region, 
whereas by explicit breaking in high-$T$ region.

Let us examine the critical exponents. Assuming the mean-field expansion, 
we expand $V$ as follows
\begin{eqnarray}
V=a_2(T)\sigma^2+a_4(T)\sigma^4+b(T)m_R\sigma+\cdots,\label{eqn:mfe}
\end{eqnarray}
where ellipsis represents higher order contributions in $m_R$ and $\sigma$ that
 is expected to be small near $T_c$ 
and the critical temperature is determined 
by the condition $a_2(T_c)=0$ for $m_R=0$. 
The three critical exponents $\beta$, $\gamma$ and $\delta$ 
are defined as follows
\begin{eqnarray}
\barq |_{m_R=0} & \simeq & 
\left(1-\frac{T}{T_c}\right)^{\beta},\label{eqn:beta}\\
\frac{\partial\barq}{\partial m_R}\bigg{|}_{m_R=0} 
& \simeq & \left(1-\frac{T}{T_c}\right)^{-\gamma},\label{eqn:gamma}\\
\barq|_{T=T_c} & \simeq & m_R^{1/\delta},\label{eqn:delta}
\end{eqnarray}
where $T<T_c$. 
Since we have confirmed that, except for $a_2(T)$, all the 
coefficients in Eq. (\ref{eqn:mfe}) do not have singular
 behavior near $T_c$ we can define 
the exponent $\gamma$, instead of Eq. (\ref{eqn:gamma}), as
\begin{eqnarray}
a_2(T) \simeq \left(1-\frac{T}{T_c}\right)^{\gamma}.
\end{eqnarray}
These critical exponents are determined numerically by the use of the 
linear log fit, for instance,
\begin{eqnarray}
\ln \barq |_{m_R=0}=\beta\ln\left(1-\frac{T}{T_c}\right)+C,
\end{eqnarray}
where $C$ is independent of $T$. In order to determine the exponent we use the 
$\chi^2$ fitting (see Figs. 7 and 8). The results are
\begin{eqnarray}
\beta \simeq 0.5~,~\gamma \simeq 1~,~\delta \simeq 3.
\end{eqnarray}
Therefore, our results confirm the Landau theory of the 
second-order phase transition, i.e., 
a second-order phase transition with the classical mean-field values for 
the critical exponents, and coincident with that obtained in Ref. \cite{HOLL}.

In order to calculate the masses of the scalar ($\sigma$) 
and pseudoscalar ($\pi$) bosons, we include 
the pseudoscalar degree of freedom in $V$ and take the second derivative. 
The values of masses are obtained by multiplying the second derivative 
by the appropriate factor $f$:
\begin{eqnarray}
M_{\sigma}^2=f\frac{\partial^2 V}{\partial \sigma^2}\Bigg{|}_{\rm min}~,~
M_{\pi}^2=f\frac{\partial^2 V}{\partial \sigma_5^2}\Bigg{|}_{\rm min}.
\end{eqnarray}
Here, ``min'' at the end of the equations means that they are evaluated 
at the minimum of $V(\sigma,\sigma_5;m_R)$. 
In this paper, we do not examine the factor $f$; rather, we study 
the ``mass ratio'' $M_{\sigma}/M_{\pi}$. 
As discussed in Refs. \cite{KUNIHIRO,BARDUCCI99}, 
below $T_c$, $M_{\pi}(T)$ weakly depends on the temperature 
and the value is dominated by the current quark mass and 
for $T \geq T_c$ the pion loses its Goldstone nature. 
On the other hand, $M_{\sigma}(T)$ decreases in association with the 
partial restoration of chiral symmetry. Therefore, above some 
temperature $T^{\ast}$, $M_{\sigma}(T)$ becomes smaller than $2M_{\pi}(T)$ 
and the width 
$\Gamma_{\sigma\rightarrow 2\pi} 
\propto \sqrt{1-4M_{\pi}^2/M_{\sigma}^2}$ vanishes. 
Figure 9 shows the temperature dependence of $M_{\sigma}/M_{\pi}$. We evaluate 
$T^{\ast}$ and find
\begin{eqnarray}
T^{\ast} \simeq 0.97 T_c.
\end{eqnarray}
If we fix the $M_\pi(T=0)$ to 140 MeV, then, $M_{\sigma}(T=0)$ turns out 
to be $668$ MeV. The temperature dependence of $M_\sigma$ and
$M_\pi$ is shown in Fig. 10.

\subsection{$T=0,\mu \neq 0$ case}
Figure 11 shows the chemical potential dependence of 
the effective potential at $T=0$ for $m_R(1\gev)=7$ MeV. 
In the chiral limit, the critical chemical potential is $\mu_c=422$ MeV. 
We note that, in $m_R(1\gev)=7$ MeV case, another extremum appears for 
relatively larger value of $\mu$. 
Figure 12 shows the $\mu$ dependence of $\sigma_{min}$. 
The quark condensate discontinuously vanishes at $\mu=\mu_c$ 
in the chiral limit, indicating the first order phase transition. 
In $m_R(1\gev)=7$ MeV case, 
$\sigma_{min}$ changes discontinuously at $\mu=\mu_c^{\ast}$, 
a little larger value than $\mu_c$. However, 
because of finite quark mass, it does not vanish for $\mu > \mu_c^{\ast}$. 
It is widely expected that as $m_R$, namely the external field, grows 
the discontinuity weakens or disappears. 
We examined up to $m_R\simeq 400$ MeV, however, the discontinuity does
not vanish.

\subsection{$T \neq 0,\mu \neq 0$ case}
Figure 13 shows the $\mu$ dependence of $\sigma_{min}$ for 
several temperatures. We expect that there is the critical end point 
$E$, where the discontinuous jump of the order parameter ends 
in the $T$-$\mu$ plane. Apparently its position is
\begin{eqnarray}
T_E \sim 100~\mbox{MeV},~\mu_E \sim 300~\mbox{MeV},
\end{eqnarray}
though it is not easy to determine accurately. The physics near the 
critical end point is interesting in the light of relativistic 
heavy-ion collision experiments \cite{CSD}. This, however, is beyond the scope
of this work.

\section{SUMMARY AND DISCUSSION}

The primary purpose of this paper was to 
study the current quark mass effects on chiral phase 
transition of QCD in the improved ladder approximation. 
We made use of the CJT effective potential because the use of 
the SDE only is not adequate for studying the 
phase transition, and moreover, it was known that, in studies of 
the SDE, there is a difficulty in removing the 
perturbative contribution from the quark condensate \cite{MIRANSKY,KUSAKA}.

To begin with, we modified the form of the CJT effective potential 
using the two representations of the 
SDE. We then studied the $t_R$ and $m_R$ 
dependence of the order parameter and obtained the reasonable results. 
Being motivated by Ref. \cite{APPELQUIST}, the critical number of massless 
flavors was also studied. Our result is consistent with lattice QCD results 
for relatively small $t_R$. 
Incidentally, our formulation of the effective potential is entirely based on 
the Higashijima-Miransky approximation. However it was known that 
the chiral Ward-Takahashi identity is broken unless gluon momentum 
squared is used as the argument of the running coupling \cite{JAIN}. 
Therefore, it is desirable to formulate the effective potential for 
the finite current quark mass independently of 
the Higashijima-Miransky approximation. 

We then extended the effective potential 
to finite temperature and density by introducing 
the functions $\dtmu$, $\stmu$ and $\ptmu$. We calculated the effective 
potential numerically and investigated current quark mass 
effects on the chiral phase structure. 
In either $T \neq 0$, $\mu=0$ or $T=0$, $\mu \neq 0$ case, 
the behavior of the order parameter in low-$T$ or low-$\mu$ region 
does not suffer the effects of 
finite, however small enough, current quark mass. 
On the other hand, in high-$T$ or high-$\mu$ region 
its behavior is perfectly dominated by the current quark mass. 
We also examined numerically 
the critical exponents $\beta$, $\gamma$ and $\delta$ and confirmed 
the Landau theory of the second-order phase transition. 
The temperature dependence of the mass ratio $M_{\sigma}/M_{\pi}$ was 
also studied. 
We found that for $T>T^{\ast}=0.97T_c$ the mass ratio 
becomes smaller than two. 
In Ref. \cite{BARDUCCI99,MARIS}, the temperature $T^{\ast}$ has been obtained 
and our results are almost coincident with theirs. 
We note that the similar temperature dependence 
of the mass ratio have been also obtained from 
the Nambu--Jona-Lasinio model \cite{KUNIHIRO} and the linear sigma 
model \cite{MATSUI}. In the previous paper \cite{KIRIYAMA2}, in the 
chiral limit, we found the tricritical point at $T_P=107$ MeV, $\mu=210$
 MeV. We found ,in $m_R(1\gev)=7$ MeV case, the position of 
the critical end point is $T_E \sim 100$ MeV, $\mu_E \sim 300$ MeV.

Finally, some comments are in order. 
The treatment of the quark and gluon propagator at finite temperature 
and density is somewhat oversimplified in the present work. 
In the future, we would like to consider 
the wave function renormalization and more 
appropriate functional form for $\stmu$, $\ptmu$ and $\dtmu$ which 
would depend on $T$ and $\mu$ explicitly. 
In particular, we should take into account the screening of 
the gluon in $\dtmu$, that is to say, the effects of the Debye (electric) mass 
and the nonperturbative magnetic mass, 
which arises from the dimensional reduction, 
$m_{mag}\sim g^2T$ at $T \neq 0$ \cite{GROSS}. 
They may affect the precise location of the critical line. 
We also note that these screening masses are probably 
related to the infrared regularization parameter $p_R$, 
though it is not easy to show it explicitly. 
Furthermore, it is well known that gauge covariance is lost, namely the 
Ward-Takahashi identity for the fermion--gauge-boson vertex is not satisfied 
when one uses the ladder approximation. Of course, gauge invariance in the 
case of the non-Abelian gauge theory is ensured by 
the Slavnov-Taylor identity which, in covariant gauges, includes a 
contribution connected with the Faddeev-Popov ghost fields. 
Thus, addressing ourselves to gauge covariance in QCD is difficult, 
however, progress has been made with the Abelian gauge theory \cite{MIRANSKY}. 
Although we suppose that results of the analysis will be essentially 
unchanged, it is preferable to use 
the correct form of the fermion--gauge-boson vertex. 
As concerns the phase structure, 
it is interesting to study the physics near the 
critical end point, for example, critical slowing down \cite{CSD}. 
We also plan to study the quark pairing including a color superconductivity 
and a ``color-flavor locking'' \cite{SCHAFER} (for $N_c=N_f=3$ case).

\acknowledgements
\section*{}
O.K. would like to thank V. A. Miransky for useful discussions and for 
reading the manuscript.

\appendix
\section*{}
In this appendix, we show the effective potential $V(\sigma,\sigma_5;m_R)$ 
explicitly. In the first place, we consider the case of zero temperature 
and finite chemical potential. 

Using Eqs. (\ref{eqn:sigma}) and (\ref{eqn:pi}), we obtain
\begin{eqnarray}
V_1 (\sigma,\sigma_5;m_R)&=&
-2\int\frac{d^4\pe}{(2\pi)^4}~\ln\frac{\stmusq+\ptmusq
+(p_4+i\mu)^2+|\vec{p}|^2}{(p_4+i\mu)^2+|\vec{p}|^2}\nonumber\\
&=& -\frac{1}{4\pi^3}\int_p
\ln\left[\frac{(\stmusq+\ptmusq+p_4^2+|\vec{p}|^2-\mu^2)^2+(2\mu p_4)^2}
{(p_4^2+|\vec{p}|^2-\mu^2)^2+(2\mu p_4)^2}\right]\nonumber\\
&&+\delta^{(1)},\label{eqn:a1}
\end{eqnarray}
where the imaginary part of $V_1$ is odd function of $p_4$; 
therefore it has been removed from Eq. (\ref{eqn:a1}) and
\begin{eqnarray}
\int_p= \int_{-\infty}^{\infty} dp_4
\int_0^{\infty}d|\vec{p}|~|\vec{p}|^2.
\end{eqnarray}
Moreover, in order to remove the divergence, 
we introduce $\delta^{(1)}=V_1 (0,0;m_R)$ 
the term independent of $\sigma$ and $\sigma_5$.

In Eq. (\ref{eqn:vcjt2}), we carry out the momentum differentiation and, then, 
use the Eqs. (\ref{eqn:dp}), (\ref{eqn:sigma}) and (\ref{eqn:pi}). 
$V_2$ is obtained as
\begin{eqnarray}
V_2 (\sigma,\sigma_5;m_R)&=&
\frac{4}{3C_2a\pi^3}\int_p\frac{(p_4^2+|\vec{p}|^2)^2
(p_4^2+|\vec{p}|^2+p_R^2)[\ln(p_4^2+|\vec{p}|^2+p_R^2)]^2}
{(p_4^2+|\vec{p}|^2+p_R^2)\ln(p_4^2+|\vec{p}|^2+p_R^2)+p_4^2+|\vec{p}|^2}
{\cal S}^2(p;\sigma,m_R)\nonumber\\
&&+\frac{4\sigma_5^2}{3\pi^3C_2a}\int_p
\frac{(p_4^2+|\vec{p}|^2)^2[\ln(p_4^2+|\vec{p}|^2+p_R^2)]^{a-2}}
{(p_4^2+|\vec{p}|^2+p_R^2)
\ln(p_4^2+|\vec{p}|^2+p_R^2)+p_4^2+|\vec{p}|^2}\nonumber\\
&&\times\frac{\left[\ln(p_4^2+|\vec{p}|^2+p_R^2)+1-a/2\right]^2}
{(p_4^2+|\vec{p}|^2+p_R^2)^3}\nonumber\\
&&+\delta^{(2)},\label{eqn:a2}
\end{eqnarray}
where the function
\begin{eqnarray}
{\cal S}(p;\sigma,m_R)&=&
-\frac{am_R}{2}\frac{\left[\ln(p_4^2+|\vec{p}|^2+p_R^2)\right]^{-a/2-1}}
{p_4^2+|\vec{p}|^2+p_R^2}\nonumber\\
&&-\frac{\sigma}{(p_4^2+|\vec{p}|^2+p_R^2)^2}
\left[\ln(p_4^2+|\vec{p}|^2+p_R^2)\right]^{a/2-2}
\left[\ln(p_4^2+|\vec{p}|^2+p_R^2)+1-\frac{a}{2}\right],\label{eqn:a3}
\end{eqnarray}
is introduced and $\delta^{(2)}=V_2 (0,0;m_R)$ is again the subtraction term 
independent of $\sigma$ and $\sigma_5$.

At finite temperature and chemical potential, the $p_4$ integration 
in Eqs. (\ref{eqn:a1}) and (\ref{eqn:a2}) is replaced by a sum over the 
fermion Matsubara frequency.

\newpage

\begin{figure}
\vskip 0.2in
\centerline{\epsfbox{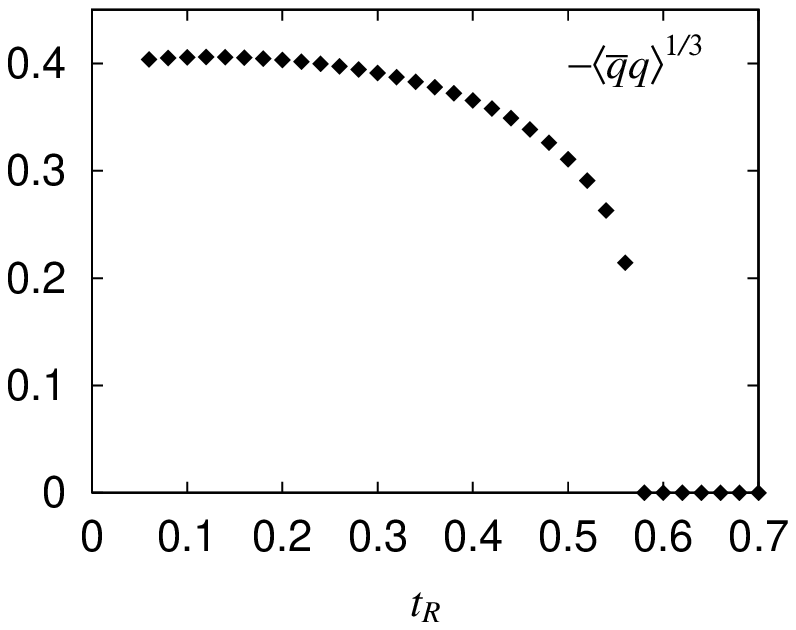}}
\vskip 0.2in
\caption{The $t_R$ dependence of $-\barq^{1/3}$ in the chiral 
limit. The critical value of $t_R$ is $(t_R)_{critical}=0.56$. 
Notice that $-\barq^{1/3}$ is taken to be dimensionless.}
\end{figure}
\vskip 0.2in

\begin{figure}
\vskip 0.2in
\centerline{\epsfbox{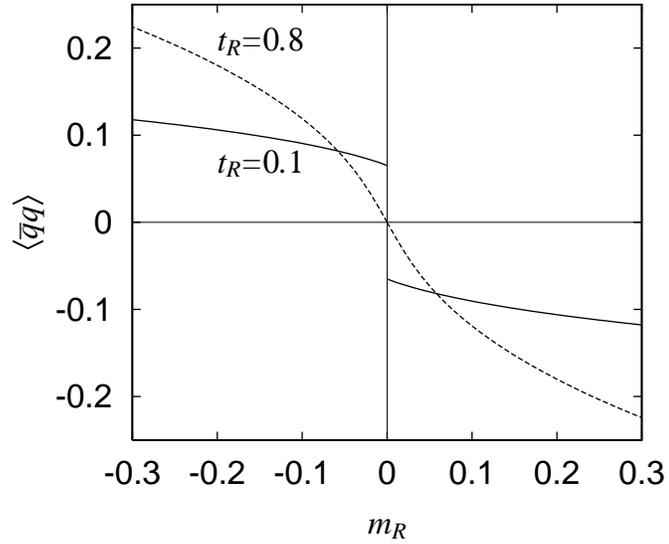}}
\vskip 0.2in
\caption{The $m_R$ dependence of $\barq$ 
for the cases $t_R=0.1$ (solid line) and $t_R=0.8$ (dotted line). 
All the quantities are taken to be dimensionless.}
\end{figure}
\vskip 0.2in

\begin{figure}
\vskip 0.2in
\epsfxsize=3.5in
\centerline{\epsfbox{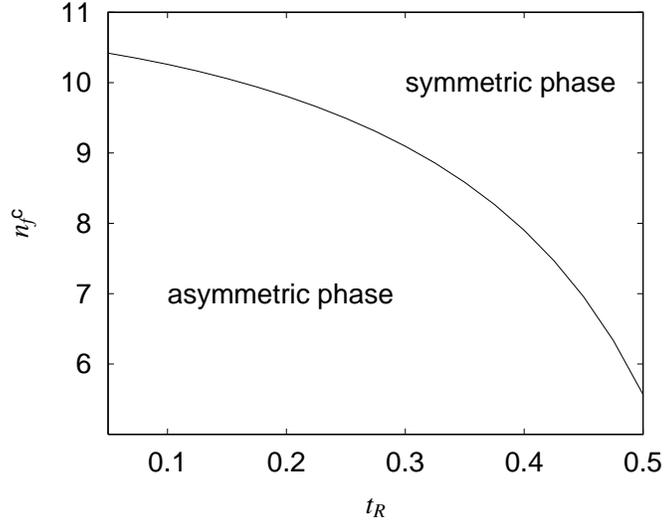}}
\vskip 0.2in
\caption{The $t_R$ dependence of $n_f^{c}$.
$n_f^{c}$ shows a tendency to decrease as $t_R$ grows.}
\end{figure}
\vskip 0.2in

\begin{figure}
\vskip 0.2in
\epsfxsize=3in
\centerline{\epsfbox{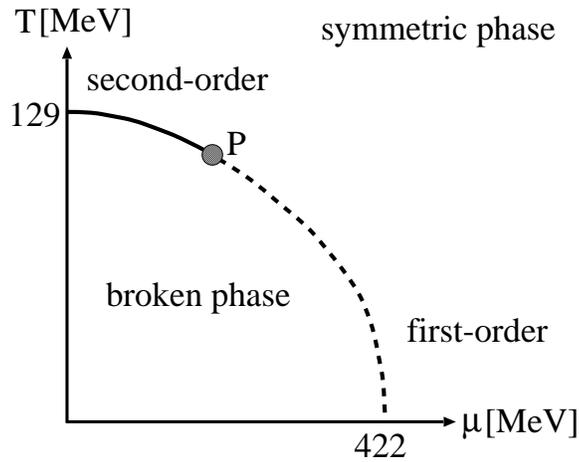}}
\vskip 0.2in
\caption{Schematic view of the phase diagram obtained from our model 
in the chiral limit [15]. 
We have found a second-order transition at $T_c=129$ MeV for $\mu=0$ and 
a first-order one at $\mu_c=422$ MeV for $T=0$. Solid line indicates the
 phase transition of second-order and dashed line indicates 
that of first-order. A tricritical point $P$ 
has been found at $T_P=107$ MeV, $\mu_P=210$ MeV.}
\end{figure}
\vskip 0.2in

\begin{figure}
\vskip 0.2in
\epsfxsize=3.5in
\centerline{\epsfbox{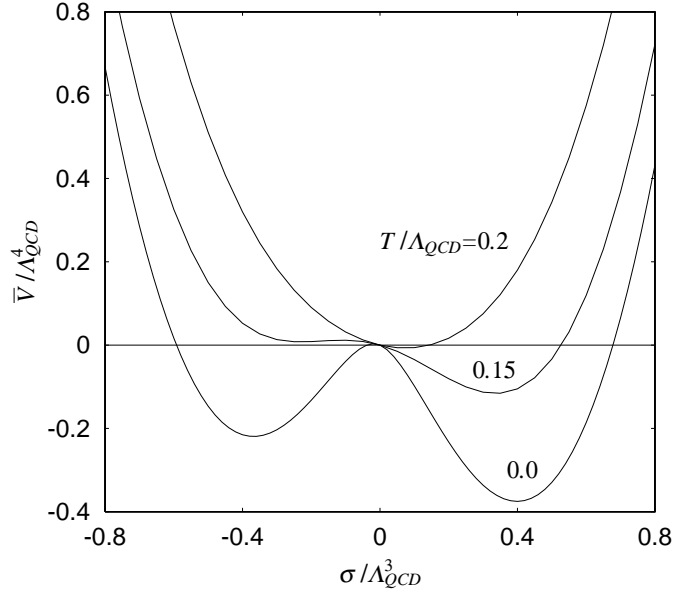}}
\vskip 0.2in
\caption{The effective potential at finite temperature and 
zero chemical potential. $\bv$ is defined by $\bv=24\pi^3V$. 
The curves show the cases $T/\Lambda_{\mt{QCD}}=0$, $0.15$ and $0.2$ 
with $m_R(1\gev)=7$ MeV.}
\end{figure}
\vskip 0.2in

\begin{figure}
\vskip 0.2in
\centerline{\epsfbox{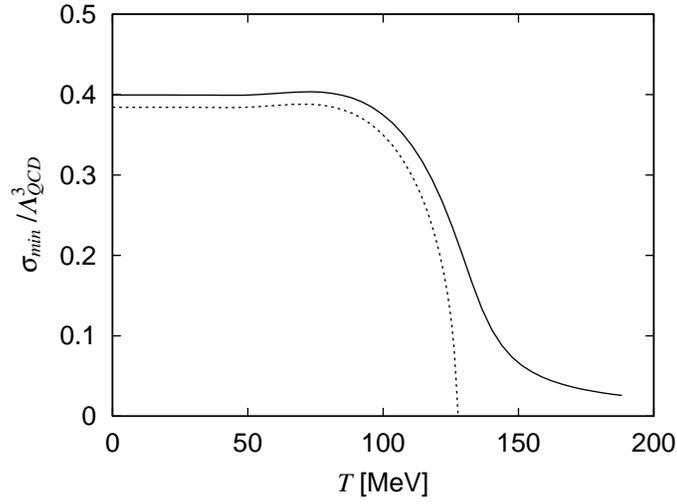}}
\vskip 0.2in
\caption{The temperature dependence of $\sigma_{min}$ at $\mu=0$. 
The curves show the cases $m_R(1\gev)=7$ MeV (solid line) 
and the chiral limit (dotted line).}
\end{figure}
\vskip 0.2in

\begin{figure}
\vskip 0.2in
\centerline{\epsfbox{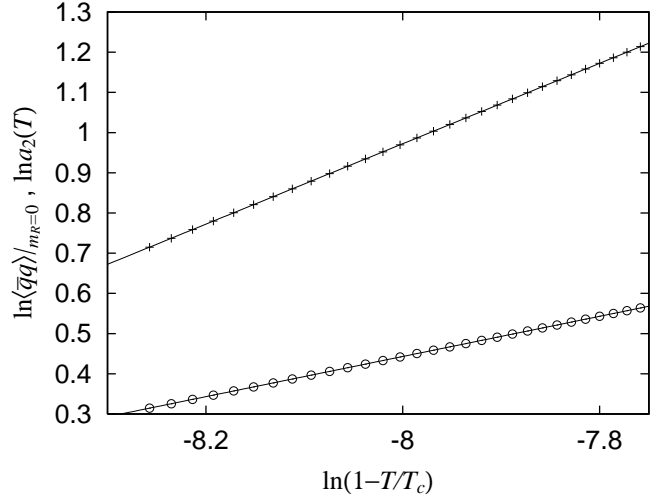}}
\vskip 0.2in
\caption{Linear log fit to $\ln\barq|_{m_R=0}$ (circle) 
and $\ln a_2(T)$ (plus). 
Here the absolute normalization of $\barq|_{m_R=0}$ and $a_2(T)$ are taken 
arbitrary. The gradient of the linear function fitted the data for 
$\ln\barq|_{m_R=0}$ and $\ln a_2(T)$ are $\beta=0.4997$ 
and $\gamma=1.000$, respectively.}
\end{figure}
\vskip 0.2in

\begin{figure}
\vskip 0.2in
\centerline{\epsfbox{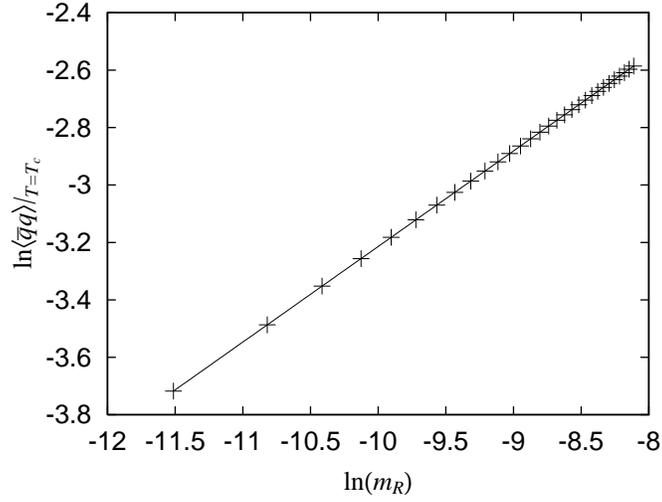}}
\vskip 0.2in
\caption{Linear log fit to $\ln\barq|_{T=T_c}$. 
The absolute normalization is taken arbitrary. The gradient of 
the linear function fitted the data is $1/\delta=0.3329$.}
\end{figure}
\vskip 0.2in

\begin{figure}
\vskip 0.2in
\centerline{\epsfbox{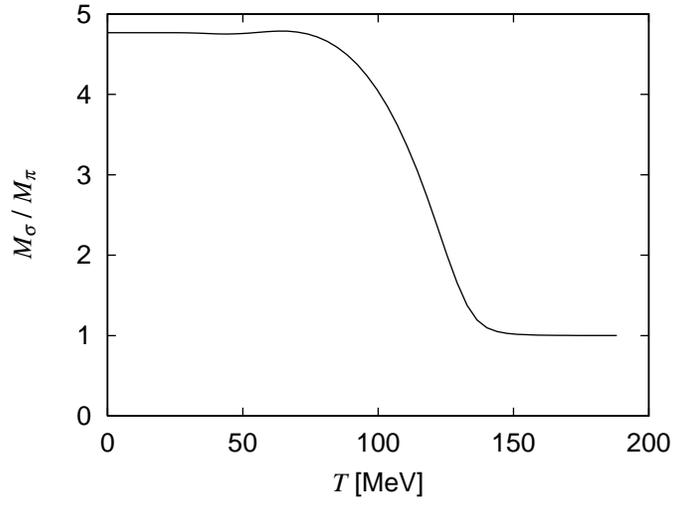}}
\vskip 0.2in
\caption{The temperature dependence of the ratio $M_{\sigma}/M_{\pi}$.}
\end{figure}
\vskip 0.2in

\begin{figure}
\vskip 0.2in
\centerline{\epsfbox{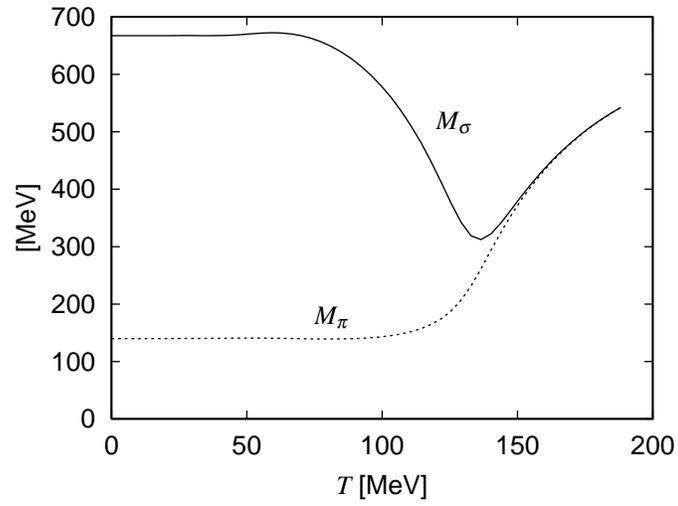}}
\vskip 0.2in
\caption{The temperature dependence of $M_{\sigma}$ (solid line) and 
$M_{\pi}$ (dotted line).}
\end{figure}
\vskip 0.2in

\begin{figure}
\vskip 0.2in
\epsfxsize=3.5in
\centerline{\epsfbox{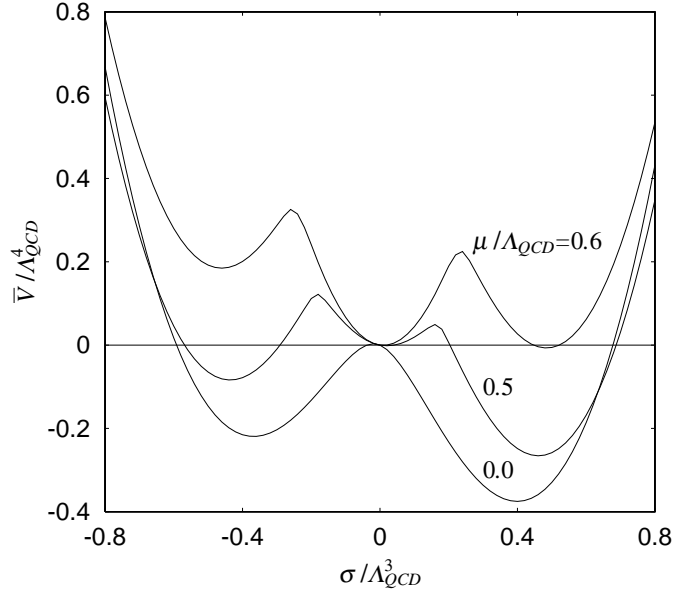}}
\vskip 0.2in
\caption{The effective potential at finite chemical potential and 
zero temperature, where $\bv=24\pi^3V$. 
The curves show the cases $\mu/\Lambda_{\mt{QCD}}=0$, $0.5$ and $0.6$ 
with $m_R(1\gev)=7$ MeV.}
\end{figure}
\vskip 0.2in

\begin{figure}
\vskip 0.2in
\centerline{\epsfbox{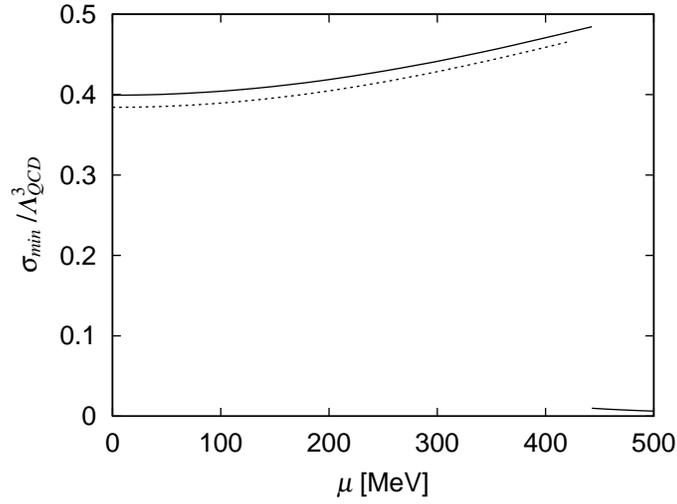}}
\vskip 0.2in
\caption{The chemical potential dependence of $\sigma_{min}$ at $T=0$. 
The curves show the cases $m_R(1\gev)=7$ MeV (solid line) 
and the chiral limit (dotted line).}
\end{figure}
\vskip 0.2in

\begin{figure}
\vskip 0.2in
\epsfxsize=4.5in
\centerline{\epsfbox{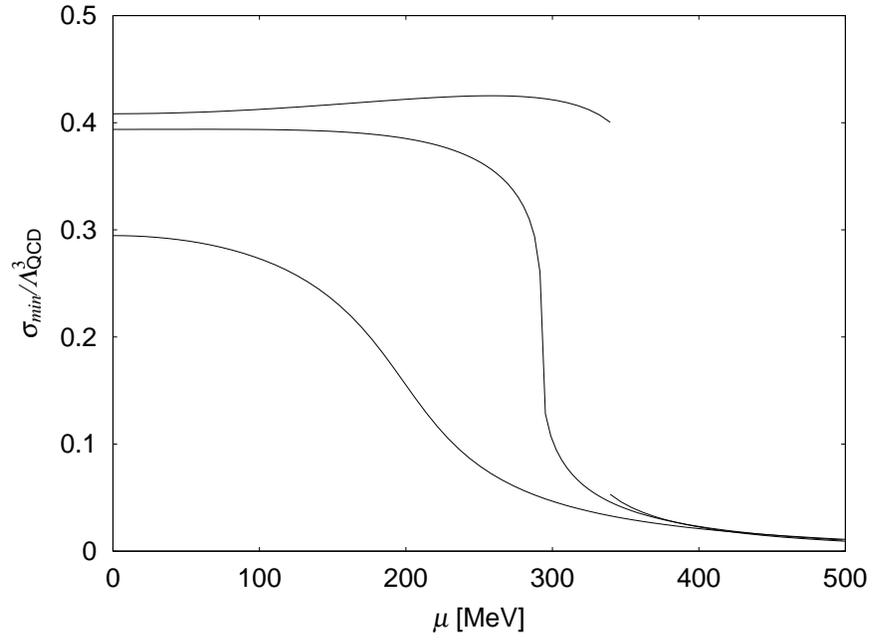}}
\vskip 0.2in
\caption{The order parameter $\sigma_{min}$ as a function of $\mu$ for
 (top to bottom) $T=80,95,120$ MeV.}
\end{figure}
\vskip 0.2in

\end{document}